\documentclass[12pt]{iopart}

\usepackage{amsfonts}

\newcommand{\beq}{\begin{equation}}
\newcommand{\eeq}{\end{equation}}
\newcommand{\beqa}{\begin{eqnarray}}
\newcommand{\eeqa}{\end{eqnarray}}
\newcommand{\beqar}{\begin{eqnarray*}}
\newcommand{\eeqar}{\end{eqnarray*}}

\newcommand{\labell}[1]{\label{#1}} %\qquad_{#1}}
\newcommand{\reef}[1]{(\ref{#1})}

\newcommand{\eg}{{\it e.g.,}\ }
\newcommand{\ie}{{\it i.e.,}\ }

\newcommand{\norm}[1]{\raise.3ex\hbox{:}#1\raise.3ex\hbox{:}}

\renewcommand{\Tr}{{\rm Tr}}
\newcommand{\STr}{{\rm STr}}

\newcommand\hi{{\rm i}}

\newcommand\prt{\partial}

\newcommand{\gsim}{\mathrel{\raisebox{-.6ex}{$\stackrel{\textstyle>}{\sim}$}}}

\newcommand{\al}{\alpha}
\newcommand{\lam}{\lambda}
\newcommand{\veps}{\varepsilon}
\newcommand\vareps{\varepsilon}

\newcommand{\eps}{\epsilon}

\newcommand{\Ga}{\Gamma}

\newcommand{\s}{\sigma}

\newcommand\cL{{\cal L}}
\newcommand\cH{{\cal H}}

\newcommand{\tG}{\widetilde{G}}

\newcommand{\phd}{\dot{\phi}}
\newcommand{\tOmega}{\widetilde{\Omega}}

\newcommand\ls{\ell_s}

\newcommand\signot{\sigma_\infty}

\renewcommand{\S}{{\rm S}}
\newcommand{\AdS}{{\rm AdS}}
\newcommand\Pp{P_\phi}

\newcommand\sig{\sigma}
\newcommand\hx{{\hat x}}

\newcommand{\bear}{\begin{eqnarray}}
\newcommand{\eear}{\end{eqnarray}}
\newcommand{\NN}{{\rm N}} %****
\newcommand\mathC{{\mathbb{C}}} %{{C\!\!\!\!C}} %{\bf C}
\newcommand\mathR{{\mathbb{R}}} %{I\!\!R}
\newcommand\mathId{{\mathbf{1}}} %{1\!\!1}
\newcommand\identity{{\mathbf{1}}}

\begin{document}

%\vspace*{0.88truein}
%\leftline{\hfill\small hep--th/0303072}\nopagebreak
%\vskip -.6ex

%\vspace*{0.48truein}

\title[Nonabelian Phenomena on D-branes]{Nonabelian Phenomena on D-branes}

\author{Robert C. Myers\footnote[1]{E-mail: {\tt rmyers@perimeterinstitute.ca}}}

\address{Perimeter Institute for Theoretical Physics,
%35 King Street North,
Waterloo, Ontario N2J 2W9 Canada}

\address{Department of Physics, McGill University,
Montr\'eal, Qu\' ebec H3A 2T8 Canada}

%\address{Department of Physics, University of Waterloo,
%Waterloo, Ontario N2L 3G1 Canada}

\begin{abstract}
A remarkable feature of D-branes is the appearance of a nonabelian
gauge theory in the description of several (nearly) coincident branes.
This nonabelian structure plays an important role in realizing various
geometric effects with D-branes. In particular, the branes' transverse
displacements are described by matrix-valued scalar fields and so
noncommutative geometry naturally appears in this framework.
I review the action governing this
nonabelian theory, as well as various related physical phenomena such
as the dielectric effect, giant gravitons and fuzzy funnels.
\end{abstract}

\maketitle

\section{Introduction}

Dirichlet branes have played a central role in all of the major advances
in string theory in the past seven years \cite{clifford}. One of the most
interesting aspects of the physics of D-branes is the appearance of a
nonabelian gauge symmetry when several D-branes are brought together.
Of course, we understand that this symmetry emerges through the appearance
of new massless states corresponding to open strings stretching between
the D-branes \cite{wite}. Thus while the number of light degrees of
freedom is proportional to N for N widely separated D-branes, this number
grows like N$^2$ for N coincident D-branes. The nonabelian symmetry
and the rapid growth in massless states are crucial elements in the
statistical mechanical entropy counting for black holes \cite{peet}, in
Maldacena's conjectured duality between type IIB superstrings in
AdS$_5\times \S^5$ and four-dimensional $\cal{N}$=4 super-Yang-Mills theory
\cite{revue}, in the development of M(atrix)-theory as a nonperturbative
description of M-theory \cite{watirev}.

Referring to the worldvolume theory for N (nearly) coincident D-branes
as a nonabelian U(N) gauge theory emphasizes the massless vector states,
which one might regard as internal excitations of the branes. Much of
the present discussion will focus on the scalar fields describing the
transverse displacements of the branes. For coincident branes, these
coordinate fields become matrix-valued appearing in the adjoint
representation of the U(N) gauge group.
These matrix-valued coordinates then provide a
natural framework where one might consider noncommutative geometry.
What has become evident after a detailed study of the world-volume action
governing the dynamics of the nonabelian U(N) theory \cite{die,wati3}
is that noncommutative geometries appear dynamically in many physical
situations in string theory.
One finds that D-branes of one dimension metamorphose into branes
of a higher dimension through noncommutative configurations.
These configurations provide hints of dualities
relating gauge theories in different dimensions, and point to a
symbiotic relationship between D-branes of all dimensions. The emerging
picture is reminiscent of the `brane democracy' speculations in very
early investigations of D-branes \cite{demo}.

One important example of this brane transmogrification is the
`dielectric effect' in which a set of D-branes are polarized
into a higher dimensional noncommutative geometry by nontrivial
background fields \cite{die}.
String theory seems to employ this brane expansion mechanism
in a variety of circumstances to regulate spacetime singularities
\cite{joemat,superstar}. So not only do string theory
and D-branes provide a natural physical framework for noncommutative
geometry, but it also seems that they may provide
a surprising realization of the old speculation that noncommutative
geometry should play a role in resolving the chaotic short-distance
structure of spacetime quantum gravity.

An outline of this paper is as follows: First, section
\ref{prelim} presents a preliminary discussion of the world-volume
actions governing the low energy physics of D-branes. Then section
\ref{noon} discusses the extension to the nonabelian action,
relevant for a system of several (nearly) coincident D-branes.
Section \ref{diel} provides an outline of the dielectric effect
for D-branes. The next section describes an important application
of the dielectric effect, namely, giant gravitons. These are
spherical D3-branes whose motion causes them to expand in an
AdS$_5\times\S^5$ background. Section \ref{bion} gives a
discussion of how noncommutative geometries can arise in the
description of intersecting branes. In \ref{horse}, we give a
short discussion of various aspects of noncommutative geometry and
fuzzy spheres which are relevant for the main text. Sections
\ref{prelim} and \ref{diel} are essentially a summary of the
material appearing in ref.~\cite{die}. Section \ref{giant}
describes the material in ref.~\cite{goliath} and  section
\ref{bion} describes that in refs.~\cite{bion} and \cite{fiv}. We
direct the interested reader to these papers for more detailed
presentations of the associated material.

\section{World-volume D-brane actions}   \label{prelim}

Within the framework of perturbative string theory,
a D$p$-brane is a
($p+1$)-dimensional extended surface in spacetime which supports
the endpoints of open strings \cite{clifford}.
The massless modes of this open string theory form a supersymmetric
U(1) gauge theory with a vector $A_a$, $9-p$ real scalars $\Phi^i$ and
their superpartner fermions --- for the most part,
the latter are ignored throughout the
following discussion. % $\Theta^\alpha$.
At leading order, the low-energy action corresponds to the
 dimensional reduction of that for ten-dimensional
U(1) super-Yang-Mills theory. However, as usual in string theory, there are
higher order $\alpha'=\ls^2$ corrections --- $\ls$ is the string length
scale. For constant field strengths, these stringy corrections can be resummed
to all orders, and the resulting action takes the Born-Infeld form \cite{bin}
\begin{equation}
S_{BI}=-T_p \int d^{p+1}\sig\
\left(e^{-\phi}\sqrt{-det(P[G+B]_{ab}+ \lambda\,F_{ab})}\right)
\labell{biact}
\end{equation}
where $T_p$ is the D$p$-brane tension %$T_p=2\pi/g/(2\pi\ls)^{p+1}$,
and $\lambda$ denotes the inverse of the (fundamental) string
tension, \ie $\lambda=2\pi\ls^2$. This Born-Infeld action
describes the couplings of the D$p$-brane to the massless
Neveu-Schwarz fields of the bulk closed string theory, \ie the
(string-frame) metric $G_{\mu\nu}$, dilaton $\phi$ and Kalb-Ramond
two-form $B_{\mu\nu}$. The interactions with the massless
Ramond-Ramond (RR) fields are incorporated in a second part of the
action, the Wess-Zumino term\cite{mike,cs,cs2}
\begin{equation}
S_{WZ}=\mu_p\int P\left[ \sum C^{(n)}\,e^B\right]e^{\lambda\,F}\ .
\labell{csact}
\end{equation}
Here $C^{(n)}$ denote the $n$-form RR potentials. Eq.~\reef{csact}
shows that a D$p$-brane is naturally charged under the
($p$+1)-form RR potential with charge $\mu_p$, and supersymmetry
dictates that $\mu_p=\pm T_p$. If we consider the special case of
the D0-brane (a point particle), the Born-Infeld action reduces to
the familiar world-line action of a point particle, where the
action is proportional to the proper length of the particle
trajectory. Actually this string theoretic D0-brane action is not
quite this simple geometric action, rather it is slightly
embellished with the additional coupling to the dilaton which
appears as a prefactor to the standard Lagrangian density. (Note,
however, that the tensors $B$ and $F$ drop out of the action since
the determinant is implicitly over a one-dimensional matrix.)
Turning to the Wess-Zumino action, we see that a D0-brane couples
to $C^{(1)}$ (a vector). Then eq.~\reef{csact} reduces to the
familiar coupling of a Maxwell field to the world-line of a point
particle, \ie
\beq \mu_0\int P\left[ C^{(1)}\right] \simeq q \int
A_\mu {dx^\mu\over d\tau}d\tau\ . \labell{simp0} \eeq

Higher dimensional D$p$-branes can also support a flux of $B+F$,
which complicates the world-volume actions above. From
eq.~\reef{csact}, we see that such a flux allows a D$p$-brane to
act as a charge source for RR potentials with a lower form degree
than $p$+1 \cite{mike}. Such configurations represent bound states
of D-branes of different dimensions\cite{wite}. To illustrate this
point, let us assume that $B$ vanishes and then we may expand the
Wess-Zumino action \reef{csact} for a D4-brane as: \beq \mu_4\int
\left( C^{(5)}+\lambda\,C^{(3)}\wedge F+
{\lambda^2\over2}C^{(1)}\wedge F\wedge F\right)\ .
\labell{csactsimp4} \eeq Hence, the D4-brane is naturally a source
for the five-form potential $C^{(5)}$. However, by introducing a
world-volume gauge field with a nontrivial first Chern class, \ie
exciting a nontrivial magnetic flux on the world-volume, the
D4-brane also sources $C^{(3)}$, which is the potential associated
with D2-branes. Hence a D4-brane with magnetic flux is naturally
interpreted as a D4-D2 bound state. Similarly a D4-brane, which
supports a gauge field with a nontrivial second Chern class, will
source the vector potential $C^{(1)}$ and is interpreted as a
D4-D0 bound state.

In both of the pieces of the D-brane action, the symbol $P[\ldots]$
denotes the pull-back of the bulk spacetime tensors to the D-brane
world-volume. Hence as already alluded to above,
the Born-Infeld action \reef{biact} has a geometric interpretation,
\ie it is
essentially the proper volume swept out by the D$p$-brane, which is
indicative of the fact that D-branes are actually dynamical objects.
This dynamics becomes more evident with an explanation of the static gauge
choice implicit in constructing the above action. To begin, we employ
spacetime diffeomorphisms to position the world-volume on a fiducial
surface defined as $x^i=0$ with $i=p+1,\ldots,9$.
With world-volume diffeomorphisms, we then match the world-volume
coordinates with the remaining spacetime coordinates on this surface,
$\sig^a=x^a$ with $a=0,1,\ldots,p$. Now the world-volume scalars $\Phi^i$
play the role of describing the transverse
displacements of the D-brane, through the identification
\begin{equation}
\labell{match}
x^i(\sig)=2\pi\ls^2\Phi^i(\sig)\qquad {\rm with}\ i=p+1,\ldots,9.
\end{equation}
With this identification the general formula for the pull-back reduces to
\bear
\labell{pull}
P[E]_{ab}&=&E_{\mu\nu}{\prt x^\mu\over\prt\sig^a}{\prt x^\nu\over\prt\sig^b}\\
&=&E_{ab}+\lambda\,E_{ai}\,\prt_{b}\Phi^i+ \lambda\,E_{ib}\,
\prt_a\Phi^i+ \lambda^2\,E_{ij}\prt_a\Phi^i\prt_b\Phi^j\ .
\nonumber \eear In this way, the expected kinetic terms for the
scalars emerge to leading order in an expansion of the Born-Infeld
action \reef{biact}. Note that our conventions are such that both
the gauge fields and world-volume scalars have the dimensions of
$length^{-1}$
--- hence the appearance of the string scale in eq.~\reef{match}.

Although it was mentioned above, we want to stress that these
world-volume actions are low energy effective actions for the
massless states of the open and closed strings, which incorporate
interactions from all disk amplitudes (all orders of tree level
for the open strings). The Born-Infeld action was originally
derived \cite{bin} using standard beta function techniques applied
to world-sheets with a boundary \cite{help}. In principle, they
could also be derived from a study of open and closed string
scattering amplitudes and it has been verified that this approach
yields the same interactions to leading order
\cite{garousi1,garousi2,igor}. As a low energy effective action
then, eqs.~\reef{biact} and \reef{csact} include an infinite
number of stringy corrections, which essentially arise through
integrating out the massive modes of the string
--- see the discussion in \cite{john}. For example, consider the Born-Infeld
action evaluated for a flat D$p$-brane in empty Minkowski space
\beqa
S_{BI}&\simeq&-T_p \int d^{p+1}x\ \sqrt{-det(\eta_{ab}+
2\pi\ls^2\,F_{ab})}
\labell{expandp}\\
&\simeq&-T_p \int d^{p+1}x\ \left(1+{(2\pi\ls^2)^2\over4} F^2
+ (2\pi\ls^2)^4 F^4 + (2\pi\ls^2)^6 F^6 + \cdots\right)
\nonumber
\eeqa
where the precise structure of the $F^4$ and $F^6$ terms may
be found in \cite{F4} and \cite{F6}, respectively. Hence as well as
the standard kinetic term for the world-volume gauge field, eq.~\reef{expandp}
includes an infinite series of higher dimension interactions, which are
suppressed as long as the typical components $\ls^2 F_{ab}$ (referred to
an orthonormal frame) are small. The square-root
expression in the first line resums this infinite series and so one may
consider arbitrary values of $\ls^2 F_{ab}$ in working with this action.
However, the full effective action would also include stringy
correction terms involving derivatives of the field strength,
\eg $\partial_aF_{bc}$ or $\partial_a\partial_bF_{cd}$ --- see, for
example, \cite{F6} or \cite{exampd}. None of these
have been incorporated in the Born-Infeld action and so one must still
demand that the variations in the field strength are relatively small,
\eg components of $\ls \partial_aF_{bc}$ are much smaller than those
of  $F_{ab}$. Of course, this discussion extends
in the obvious way to derivatives of the scalar fields $\Phi^i$.

At this point, we should also note that the bulk supergravity fields
appearing in eqs.~\reef{biact} and \reef{csact} are
in general functions of all of the spacetime coordinates, and so
they are implicitly functionals of the world-volume scalars.
In static gauge, the bulk fields are evaluated in terms of a
Taylor series expansion around the fiducial surface $x^i=0$.
For example, the metric functional appearing in
the D-brane action would be given by
\bear
G_{\mu\nu}&=&\exp\left[\lambda\Phi^i\,{\prt_{x^i}}\right]G^0_{\mu\nu}
(\sigma^a,x^i)|_{x^i=0}
\labell{slick}\\
&=&\sum_{n=0}^\infty {\lambda^n\over
n!}\,\Phi^{i_1}\cdots\Phi^{i_n}\,
(\prt_{x^{i_1}}\cdots\prt_{x^{i_n}})G^0_{\mu\nu}
(\sigma^a,x^i)|_{x^i=0}\ . \nonumber \eear
Hence the world-volume action implicitly incorporates an infinite
class of higher dimension interactions involving derivatives of
the bulk fields as well. However, beyond this class of
interactions incorporated in eqs.~\reef{biact} and \reef{csact},
once again the full effective action includes other higher
derivative bulk field corrections \cite{cs2,more2,more1, more3}.
It is probably fair to say that the precise domain of validity of
the D-brane action from the point of view of the bulk fields is
poorly understood.

\section{Nonabelian D-brane action} \label{noon}

As N parallel D-branes approach each other, the ground state modes of
strings stretching between the different D-branes become massless. These
extra massless states carry the appropriate charges to fill out
representations under a U(N) symmetry. Hence the U(1)$^{\rm N}$ of the
individual D-branes is enhanced to the nonabelian group U(N) for the
coincident D-branes \cite{wite}. The vector $A_a$ becomes a nonabelian
gauge field
\begin{equation}
A_a=A_a^{(n)}T_n,
\qquad F_{ab}=\prt_a A_b-\prt_b A_a+i[A_a,A_b]
\labell{gauge}
\end{equation}
where $T_{n}$ are $\NN^2$ hermitian generators
with $\Tr(T_{n}\,T_m)=\NN\,\delta_{n m}$.
Central to the following is that the scalars $\Phi^i$ are
also matrix-valued transforming in the adjoint of U(N).
The covariant derivative of the scalar fields is given by
\begin{equation}
D_a\Phi^i=\prt_a\Phi^i+i[A_a,\Phi^i]\ .
\labell{scalder}
\end{equation}

Understanding how to accommodate this U(N) gauge symmetry in the
world-volume action is an interesting puzzle. For example, the
geometric meaning (or even the validity) of eqs.~\reef{match} or
\reef{slick} seems uncertain when the scalars on the right hand
side are matrix-valued. In fact, the identification of the scalars
as transverse displacements of the branes does remain roughly
correct. Some intuition comes from the case where the scalars are
commuting matrices and the gauge symmetry can be used to
simultaneously diagonalize all of them. In this case, one
interpretes the N eigenvalues of the diagonal $\Phi^i$ as
representing the displacements of the N constituent D-branes ---
see, \eg \cite{watirev}. Further the gauge symmetry may be used to
simultaneously interchange any pair of eigenvalues in each of the
scalars and so ensures that the branes are indistinguishable. Of
course, to describe noncommutative geometries, we will be more
interested in the case where the scalars do not commute and so
cannot be simultaneously diagonlized.

Refs.~\cite{die} and \cite{wati3} made progress in constructing the
world-volume
action describing the dynamics of nonabelian D-branes. The essential strategy
in both of these papers was to
construct an action which was consistent with the familiar string theory
symmetry of T-duality \cite{tdual}. Acting on D-branes, T-duality acts to
change the dimension
of the world-volume \cite{clifford}. The two possibilities are:
{\it (i)} if a coordinate transverse to the D$p$-brane, \eg $y=x^{p+1}$, is
T-dualized, it becomes a D$(p+1)$-brane where $y$ is now the
extra world-volume direction; and {\it (ii)} if a world-volume coordinate
on the D$p$-brane, \eg $y=x^{p}$, is
T-dualized, it becomes a D$(p-1)$-brane where $y$ is now an
extra transverse direction. Under these transformations, the role of the
corresponding world-volume fields change as
\begin{equation}
(i)\ \Phi^{p+1}\,\rightarrow\, A_{p+1}\ ,
\qquad\qquad
(ii)\ A_p\,\rightarrow\,\Phi^p\ ,
\labell{rule1}
\end{equation}
while the remaining components of $A$ and scalars $\Phi$ are left
unchanged. Hence in constructing the nonabelian action, one can begin
with the D9-brane theory, which contains no scalars since the world-volume
fills the entire spacetime. In this case, the nonabelian extension of
eqs.~\reef{biact} and \reef{csact} is given by simply introducing an overall
trace over gauge indices of the nonabelian field strengths appearing in the
action \cite{moremike}. Then applying T-duality transformations on $9-p$
directions yields
the nonabelian action for a D$p$-brane. Of course, in this construction, one
also T-dualizes the background supergravity fields according to the known
transformation rules \cite{tdual,Tdual,Tdual1}.
As in the abelian theory, the result for nonabelian action
has two distinct pieces \cite{die,wati3}: the Born-Infeld term
\bear
{S}_{BI}&=&-T_p \int d^{p+1}\sigma\,\STr\left(e^{-\phi}\, \sqrt{\det(Q^i{}_j)}
\right.\nonumber\\
&&\left.\qquad\qquad \times\ \sqrt{-\det\left(
P\left[E_{ab}+E_{ai}(Q^{-1}-\delta)^{ij}E_{jb}\right]+
\lambda\,F_{ab}\right)} \right), \labell{finalbi} \eear with
%\begin{equation}
$E_{\mu\nu}=G_{\mu\nu}+B_{\mu\nu}$
%\qquad{\rm and}\qquad
and $Q^i{}_j\equiv\delta^i{}_j+i\lambda\,[\Phi^i,\Phi^k]\,E_{kj}$;
%\labell{extra6}
%\end{equation}
and the Wess-Zumino term
\begin{equation}
S_{WZ}=\mu_p\int \STr\left(P\left[e^{i\lambda\,\hi_\Phi \hi_\Phi}
( \sum C^{(n)}\,e^B)\right] e^{\lambda\,F}\right)\ .
\labell{finalcs}
\end{equation}

Let us enumerate the nonabelian features of this action:
%\begin{enumerate}
%\item
\vskip 1ex
\noindent 1.~\textit{Nonabelian field strength}:
The $F_{ab}$ appearing explicitly in both terms
is now nonabelian, of course.
%\item
\vskip 1ex
\noindent 2.~\textit{Nonabelian Taylor expansion}:
The bulk supergravity fields
are again functions of all of the spacetime coordinates, and so
they are implicitly functionals of the nonabelian scalars.
In the action (\ref{finalbi},\ref{finalcs}), these bulk fields are
again interpreted in terms of a Taylor expansion as in eq.~\reef{slick},
however the transverse displacements are now matrix-valued.
the D-brane action would be given by a {nonabelian} Taylor expansion

%\item
\vskip 1ex
\noindent 3.~\textit{Nonabelian Pullback}:
As was noted in refs.~\cite{hull,dorn}, the pullback of various spacetime tensors
to the world-volume must now involve covariant derivatives of the nonabelian scalars
in order to be consistent with the U(N) gauge symmetry. Hence eq.~\reef{pull} is
replaced by
\begin{equation}
P[E]_{ab}=E_{ab}+\lambda\,E_{ai}\,D_{b}\Phi^i+ \lambda\,E_{ib}\,
D_a\Phi^i+ \lambda^2\,E_{ij}D_a\Phi^iD_b\Phi^j\ . \labell{pulla}
\end{equation}
%\item
%\vskip 1ex
\noindent 4.~\textit{Nonabelian Interior Product}:
In the Wess-Zumino term \reef{finalcs},
$\hi_\Phi$ denotes the interior product with $\Phi^i$ regarded
as a vector in the transverse space, {\it e.g.}, acting on an $n$-form
$C^{(n)}={1\over n!}C^{(n)}_{\mu_1\cdots\mu_n} dx^{\mu_1}\cdots dx^{\mu_n}$,
we have
\begin{equation}
\hi_\Phi\hi_\Phi C^{(n)}={1\over2(n-2)!}\,[\Phi^i,\Phi^j]\,
C^{(n)}_{ji\mu_3\cdots\mu_n}dx^{\mu_3}\cdots dx^{\mu_n}
\ . \labell{inside}
\end{equation}
Note that acting on forms, the interior product is an anticommuting operator
and hence for an ordinary vector (\ie a vector $v^i$ with values in
$\mathR^{9-p}$):
$\hi_v\hi_vC^{(n)}=0$. It is only because the scalars $\Phi$ are matrix-valued
that eq.~\reef{inside} yields a nontrivial result.
%\item
\vskip 1ex
\noindent 5.~\textit{Nonabelian Gauge Trace}:
As is evident above, both parts of the action
are highly nonlinear functionals of the nonabelian fields, and so
eqs.~\reef{finalbi}
and \reef{finalcs} would be incomplete without a precise definition for the
ordering of these fields under the gauge trace. Above, $\STr$ denotes the
maximally symmetric trace \cite{arkady}. To be precise, the trace includes a
symmetric average over all orderings of $F_{ab}$,
$D_a\Phi^i$, $[\Phi^i,\Phi^j]$ and the individual $\Phi^k$ appearing in the
nonabelian Taylor expansions of the background fields. This choice matches
that inferred from Matrix theory \cite{wati4}, and a similar symmetrization
arises in the leading order analysis of the boundary $\beta$ functions
\cite{dorn}. However, we should note that
with this definition an expansion of the Born-Infeld term \reef{biact} does
agree with the string theory to fourth order in $F$ \cite{arkady,ark2},
but it does not seem to capture the full physics of the nonabelian
fields in the infrared limit at higher orders \cite{wati5}  --- we expand on
this point below.
%\end{enumerate}
\vskip 1ex

Some general comments on the nonabelian action are as follows:
In the Born-Infeld term \reef{finalbi}, there are now two determinant factors
as compared to one in the abelian action \reef{biact}. The second determinant
in eq.~\reef{finalbi} is a slightly modified version of that in
eq.~\reef{biact}. One might think of this as the kinetic factor, since to
leading order in the low
energy expansion, it yields the familiar kinetic terms for the gauge field
and scalars.
In the same way, one can think of the new first factor as the potential
factor, since to leading order in the low energy expansion, it reproduces the
nonabelian scalar potential expected for the super-Yang-Mills theory --- see
eq.~\reef{trivpot} below. Further note that the first factor reduces
to simply one when the scalar fields are commuting, even
for general background fields.

The form of the action and in particular the functional dependence
of the bulk fields on the adjoint scalars can be verified
in a number of independent ways. Douglas \cite{moremike} observed
on general grounds whatever their form, the
nonabelian world-volume actions should contain a single gauge trace,
as do both eqs.~\reef{finalbi} and \reef{finalcs}. This observation
stems from the fact that the action should encode only the
low energy interactions derivable from disk amplitudes in superstring
theory. Since the disk has a single boundary,
the single gauge trace arises from the standard open
string prescription of tracing over Chan-Paton factors on each
world-sheet boundary. Further then one may note \cite{moremike}
that the only difference in the
superstring amplitudes between the U(1) and the U(N) theories is that
the amplitudes in the latter case are multiplied by an additional trace
of Chan-Paton factors. Hence up to commutator `corrections',
the low energy interactions should be the same in both cases. Hence
since the background fields are functionals of the neutral U(1) scalars
in the abelian theory, they must be precisely the same functionals
of the adjoint scalars in the nonabelian theory, up to commutator
corrections. The interactions involving the nonabelian inner product
in the Wess-Zumino action \reef{finalcs} provide one class of
commutator corrections. The functional dependence on the adjoint scalars
also agrees with the linearized couplings for the bulk fields
derived from Matrix theory \cite{wati4}. As an aside, we would add
that the technology developed in Matrix theory remains useful in
gaining intuition and manipulating these nonabelian functionals
\cite{host,hiroshi}. Finally we add that by the direct examination of
string scattering amplitudes
using the methods of refs.~\cite{garousi1} and \cite{igor}, one can
verify at low orders the form of the nonabelian interactions
in eqs.~\reef{finalbi} and \reef{finalcs}, including
the appearance of the new commutator interactions in the nonabelian
Wess-Zumino action \cite{garousi3}.

As noted above, the symmetric trace perscription is known {\it not} to
agree with the full effective string action \cite{wati5}.
Rather at sixth order and higher in the world-volume field strength,
additional terms involving commutators of field strengths must be added
to the action \cite{bain}. The source of this shortcoming is clear.
Recall from the discussion following eq.~\reef{expandp} that in the
abelian action we have disgarded all interactions involving derivatives
of the field strength. However, this prescription is ambiguous in the
nonabelian theory since
\beq
[D_a,D_b]F_{cd}=i[F_{ab},F_{cd}]\ .
\labell{amcom}
\eeq
It is clear that with the symmetric trace we have eliminated all
derivative terms, including those antisymmetric combinations that
might contribute commutators of field strengths. One might choose to
improve the action by reinstating these commutators. This problem has
been extensively studied and the commutator corrections at order $F^6$
are known \cite{bain}. Considering the supersymmetric extension of the
nonabelian action \cite{lotscom,F6} has lead to a powerful iterative
technique relying on stable holomorphic bundles \cite{iterate1} which
seems to provide a constructive approach to determine the entire effective
open string action, including all higher derivative terms and fermion
contributions as well. A similar iterative procedure seems to emerge
from study the Seiberg-Witten map \cite{seiwit}
in the context of a noncommutative world-volume theory \cite{iterate2}.

As described below eq.~\reef{csact}, an individual D$p$-brane
couples not only to the RR potential with form degree $n=p+1$, but
also to the RR potentials with $n=p-1,p-3,\ldots$ through the
exponentials of $B$ and $F$ appearing in the Wess-Zumino action
\reef{csact}. Above in eq.~\reef{finalcs}, $\hi_\Phi\hi_\Phi$ is
an operator of form degree --2, and so world-volume interactions
appear in the nonabelian action \reef{finalcs} involving the
higher RR forms. Hence in the nonabelian theory, a D$p$-brane can
also couple to the RR potentials with $n=p+3,p+5,\ldots$ through
the additional commutator interactions. To make these couplings
more explicit, consider the D0-brane action (for which $F$
vanishes): \bear S_{CS}&=&\mu_0\int \STr\left(P\left[C^{(1)}+
i\lambda\,\hi_\Phi \hi_\Phi\left(C^{(3)}+C^{(1)}B\right)
\vphantom{\lambda^4\over24}\right.\right.
\nonumber\\
&&\quad \qquad-{\lambda^2\over2}(\hi_\Phi \hi_\Phi)^2\left(C^{(5)}
+C^{(3)}B+{1\over2}C^{(1)}B^2\right)
\nonumber\\
&&\quad -i{\lambda^3\over6}(\hi_\Phi \hi_\Phi)^3\left(
C^{(7)}+C^{(5)}B +{1\over2}C^{(3)}B^2+{1\over6}C^{(1)}B^3\right)
\labell{cszero}\\
&&\quad \left.\left.+{\lambda^4\over24}(\hi_\Phi \hi_\Phi)^4\left(
C^{(9)}+C^{(7)}B+{1\over2}C^{(5)}B^2
+{1\over6}C^{(3)}B^3+{1\over24}C^{(1)}B^4\right)\right]\right).
\nonumber \eear Of course, these interactions are reminiscent of
those appearing in Matrix theory \cite{matrix,matbrane}. For
example, eq.~\reef{cszero} includes a linear coupling to
$C^{(3)}$, which is the potential corresponding to D2-brane
charge, \bear &&i\lambda\,\mu_0\int \Tr\, P\left[\hi_\Phi \hi_\Phi
C^{(3)}\right]
\nonumber\\
&&\qquad\quad =i{\lambda\over2}\mu_0\int dt\ \Tr
\left(C_{tjk}^{(3)}(\Phi,t)\,[\Phi^k,\Phi^j] +\lambda
C^{(3)}_{ijk}(\Phi,t)\,D_t\Phi^k\,[\Phi^k,\Phi^j] \right)
\labell{magic} \eear where we assume that $\sigma^0=t$ in static
gauge. Note that the first term on the right hand side has the
form of a source for D2-brane charge. This is essentially the
interaction central to the construction of D2-branes in Matrix
theory with the large N limit \cite{matrix,matbrane}. Here,
however, with finite N, this term would vanish  upon taking the
trace  if $C_{tjk}^{(3)}$ was simply  a function of the
world-volume coordinate $t$ (since $[\Phi^k,\Phi^j] \in{\rm
SU(N)}\,)$. However, in general these three-form components  are
functionals of $\Phi^i$. Hence, while there would be no `monopole'
coupling to D2-brane charge, nontrivial expectation values of the
scalars can give rise to couplings to an infinite series of higher
`multipole' moments \cite{host}.

\section{Dielectric Branes}   \label{diel}

In this section, we wish to consider certain physical
effects arising from the new nonabelian interactions
in the world-volume action, given by eqs.~\reef{finalbi} and \reef{finalcs}.
To begin,
consider the scalar potential for D$p$-branes in flat space, \ie
$G_{\mu\nu}=\eta_{\mu\nu}$ with all other fields vanishing.
In this case, the entire scalar potential originates in the Born-Infeld
term \reef{finalbi} as
\begin{equation}
V=T_p\,\Tr\sqrt{det(Q^i{}_j)}= \NN T_p-{T_p\lambda^2\over4}
\Tr([\Phi^i,\Phi^j]\,[\Phi^i,\Phi^j])+\ldots \labell{trivpot}
%&&\qquad\qquad-{T_p\lambda^4\over8}
%\Tr([\Phi^i,\Phi^j]\,[\Phi^j,\Phi^k]\,[\Phi^k,\Phi^l]\,[\Phi^l,\Phi^i])
%+{T_p\lambda^4\over32}\Tr(([\Phi^i,\Phi^j]\,[\Phi^j,\Phi^i])^2)
%+\cdots
%\nonumber
\end{equation}
The commutator-squared term
corresponds to the potential for ten-dimensional U(N)
super-Yang-Mills theory reduced to $p+1$ dimensions.
A nontrivial set of extrema of this potential is given by taking the $9-p$ scalars as
constant commuting matrices, \ie
\begin{equation}
\labell{commat}
[\Phi^i,\Phi^j]=0
\end{equation}
for all $i$ and $j$. Since they are commuting, the $\Phi^i$ may
be simultaneously diagonalized and as discussed above,
the eigenvalues are interpreted as the separated positions of N fundamental D$p$-branes
in the transverse space. This solution reflects the fact that a system of
N parallel D$p$-branes is supersymmetric, and so they can sit in
static equilibrium with arbitrary separations in the transverse space
\cite{clifford}.

From the results described in the previous section,
it is clear that in going from flat space to general background
fields, the scalar potential is modified by new interactions and so one
should reconsider the analysis of the extrema.
It turns out that this yields an interesting physical effect
that is a precise analog for D-branes of the dielectric effect in ordinary electromagnetism.
That is when D$p$-branes are placed  in a nontrivial background
field for which the D$p$-branes would normally be regarded as neutral,
\eg nontrivial $F^{(n)}$ with $n>p+2$, new terms will be induced
in the scalar potential, and generically one should expect that there
will be new extrema beyond those found in flat space, \ie eq.~\reef{commat}.
In particular, there can be nontrivial extrema
with noncommuting expectation values of the $\Phi^i$, \eg with
$\Tr\Phi^i=0$ but $\Tr(\Phi^i)^2\ne0$. This would correspond to
the external fields `polarizing' the D$p$-branes to expand
into a (higher dimensional)
noncommutative world-volume geometry. This is the analog of the familiar
electromagnetic process where an external field may induce
a separation of charges in neutral materials. In this case, the
polarized material will then carry an electric dipole (and possibly
higher multipoles). The latter is also seen in the D-brane analog. When
the world-volume theory is at a noncommutative extremum, the
gauge traces of products of scalars will be nonvanishing in various interactions
involving the supergravity fields. Hence at such an extremum, the
D$p$-branes act as sources for the latter bulk fields.

To make these ideas explicit, we will now illustrate the
process with a simple example. We consider N D0-branes
in a constant background RR field $F^{(4)}$, \ie the field strength
associated with D2-brane charge. We find that the D0-branes expand into
a noncommutative two-sphere which represents a spherical bound state
of a D2-brane and N D0-branes.

Consider a background where only
RR four-form field strength is nonvanishing with
\begin{equation}
F^{(4)}_{tijk}=-2f \vareps_{ijk}\qquad{\rm for}\ i,j,k\in
\lbrace 1,2,3\rbrace
\labell{backg}
\end{equation}
with $f$ a constant (of dimensions $length^{-1}$).
Since $F^{(4)}=dC^{(3)}$,
we must consider the coupling of the D0-branes
to the RR three-form potential, which is given above in eq.~\reef{magic}.
If one explicitly introduces the nonabelian Taylor expansion \reef{slick},
one finds the leading order interaction may be written as
\begin{equation}
{i\over3}\lambda^2\mu_0\int dt\,\Tr\left(\Phi^i\Phi^j\Phi^k\right)
F^{(4)}_{tijk}(t)\ .
\labell{interact3}
\end{equation}
This final form might have been anticipated
since one should expect that the world-volume potential can only depend on
gauge invariant expressions of the background field. Given that
we are considering a constant background $F^{(4)}$,
the higher order terms implicit in eq.~\reef{magic} will
vanish as they can only involve spacetime derivatives of the four-form
field strength. Combining eq.~\reef{interact3} with
the leading order Born-Infeld potential \reef{trivpot} yields the
scalar potential of interest for the present problem
\begin{equation}
V(\Phi)=\NN T_0-{\lambda^2T_0\over4}\Tr([\Phi^i,\Phi^j]^2)
-{i\over3}\lambda^2\mu_0\Tr\left(\Phi^i\Phi^j\Phi^k\right)
F^{(4)}_{tijk}(t)\ .
\labell{potential}
\end{equation}

Substituting in the (static) background field \reef{backg} and $\mu_0=T_0$,
$\delta V(\Phi)/\delta\Phi^i=0$ yields
\begin{equation}
0=[[\Phi^i,\Phi^j],\Phi^j]+{i}\,f\vareps_{ijk}[\Phi^j,\Phi^k]\ .
\labell{eqmot}
\end{equation}
Note that commuting matrices \reef{commat} describing separated D0-branes
still solve this equation.
The value of the potential for these solutions is simply $V_0=\NN T_0$,
the mass of N D0-branes. Another interesting solution of eq.~\reef{eqmot} is
\begin{equation}
\Phi^i={f\over2}\,\alpha^i
\labell{solu1}
\end{equation}
where $\alpha^i$ are any N$\times$N matrix representation of
the SU(2) algebra
\begin{equation}
[\al^i,\al^j]=2i\,\vareps_{ijk}\,\al^k\ .
\labell{su2}
\end{equation}
For the moment, let us focus on the irreducible representation for
which one finds
\begin{equation}
\Tr[(\al^i)^2]={\NN\over3}(\NN^2-1) \quad{\rm for}\ i=1,2,3.
\labell{trace}
\end{equation}
Now evaluating the value of the potential \reef{potential}
for this new solution yields
\begin{equation}
V_\NN=\NN T_0-{T_0\lambda^2f^2\over6}\sum_{i=1}^3\Tr[(\Phi^i)^2]
=\NN T_0-{\pi^2\ls^3f^4\over6g}\NN^3\left(1-{1\over\NN^2}\right)
\labell{evpot}
\end{equation}
using $T_0=1/(g\ls)$.
Hence the noncommutative solution \reef{solu1} has lower energy
than a solution of commuting matrices, and so the latter
configuration of separated D0-branes is unstable towards
condensing out into this noncommutative solution.
One can also consider reducible representations of the
SU(2) algebra \reef{su2}, however, one finds that the corresponding energy
is always larger than that in eq.~\reef{evpot}. Hence it seems that
the irreducible representation describes the ground state of the system.

Geometrically, one can recognize the SU(2) algebra
as that corresponding to the noncommutative or fuzzy two-sphere
\cite{hoppe,fuzz}. The physical size of
the fuzzy two-sphere is given by
\begin{equation}
R=\lambda\left(\sum_{i=1}^3\Tr[(\Phi^i)^2]/\NN\right)^{1/2}=\
\pi\ls^2f\NN\left(1-{1\over \NN^2}\right)^{1/2} \labell{radius}
\end{equation}
in the ground state solution.
From the Matrix theory construction of Kabat
and Taylor \cite{wati2}, one can infer this ground state is not simply
a spherical arrangement of D0-branes rather the noncommutative solution actually
represents a spherical D2-brane with N D0-branes bound to it.
In the present context, the latter can be verified by seeing that this
configuration has a `dipole' coupling to the RR four-form.
The precise form of this
coupling is calculated by substituting the noncommutative
scalar solution \reef{solu1} into the world-volume interaction
\reef{interact3}, which yields
\begin{equation}
-{R^3\over3\pi g\ls^3}\left(1-{1\over \NN^2}\right)^{-1/2}\int dt\,F^{(4)}_{t123}\ .
\labell{dipole}
\end{equation}
for the ground state solution. Physically this $F^{(4)}$-dipole moment arises
because antipodal surface elements on the sphere have the opposite
orientation and so form small pairs of separated
membranes and anti-membranes. Of course, the spherical configuration
carries no net D2-brane charge.

Given that the noncommutative ground state solution corresponds to a bound
state of a spherical D2-brane and N D0-branes, one might attempt to match
the above results using the dual formulation. That is, this system can
be analyzed
from the point of view of the (abelian) world-volume theory of a D2-brane.
In this case, one would consider a spherical D2-brane carrying a
flux of the U(1) gauge field strength representing the N bound D0-branes,
and at the same time, sitting in the background of the constant RR four-form
field strength \reef{backg}.
In fact, one does find stable static solutions, but what is more
surprising is how well the results match those calculated in the framework
of the D0-branes. The results for the energy, radius and
dipole coupling are the same as in eqs.~\reef{evpot}, \reef{radius}
and \reef{dipole}, respectively, except that the factors of $(1-1/\NN^2)$
are absent \cite{die}. Hence for large N, the two calculations agree up to
$1/\NN^2$ corrections.

One expects that the D2-brane calculations would be valid when $R\gg\ls$
while naively the D0-brane calculations would be valid when $R\ll\ls$.
Hence it appears there is no common domain where the two pictures can
both produce reliable results. However, a more careful consideration
of range of validity of the D0-brane calculations only requires that
$R\ll\sqrt{\NN}\ls$. This estimate is found by requiring that the
scalar field commutators appearing in the full nonabelian potential
\reef{trivpot} are small so that the Taylor expansion of the square root
converges rapidly. Hence for large N, there is a large domain of
overlap where both of the dual pictures are reliable.
Note the density of D0-branes on the two-sphere is
$\NN/(4\pi R^2)$. However, even if $R$ is macroscopic it is still bounded by
$R\ll\sqrt{\NN}\ls$ and so this density must
be large compared to the string scale, \ie the density is much larger than
$1/\ls^2$. With such large densities, one can imagine the discreteness of the fuzzy
sphere is essentially lost and so there is good agreement with the continuum
sphere of the D2-brane picture. More discussion on the noncommutative
geometry appears in \ref{horse}.

Finally note that the Born-Infeld action  contains couplings to the
Neveu-Schwarz two-form which are similar to that in eq.~\reef{interact3}.
{}From the expansion of $\sqrt{det(Q)}$, one finds a cubic interaction
\begin{equation}
{i\over3}\lambda^2T_0\int dt\,\Tr\left(\Phi^i\Phi^j\Phi^k\right)
H_{ijk}(t)\ .
\labell{nese3}
\end{equation}
Hence the noncommutative ground state,
which has $\Tr\left(\Phi^i\Phi^j\Phi^k\right)\ne0$, also
acts as a source of the $B$ field with
\begin{equation}
-{R_0^3\over3\pi g\ls^3}\left(1-{1\over \NN^2}\right)\int dt\,H_{123}\ .
\labell{hdipole}
\end{equation}
This coupling is perhaps not so surprising given that the
noncommutative ground state represents the bound state of a
spherical D2-brane and N D0-branes. Explicit supergravity
solutions describing D2-D0 bound states with a planar geometry
have been found \cite{useful}, and are known to carry a long-range
$H$ field with the same profile as the RR field strength
$F^{(4)}$. One can also derive this coupling from the dual
D2-brane formulation. Furthermore, we observe that the presence of
this coupling \reef{nese3} means that we would find an analogous
dielectric effect if the N D0-branes were placed in a constant
background $H$ field. This mechanism plays a role in describing
D-branes in the spacetime background corresponding to a WZW model
\cite{WZW1,WZW3}. It seems that quantum group symmetries may be
useful in understanding these noncommutative configurations
\cite{qWZW}.

The example considered above must be considered simply a toy
calculation demonstrating the essential features of the dielectric
effect for D-branes. A more complete calculation would require
analyzing the D0-branes in a consistent supergravity background.
For example, the present case could be extended to consider the
asymptotic supergravity fields of a D2-brane, where the RR
four-form would be slowly varying but the metric and dilaton
fields would also be nontrivial. Alternatively, one can find
solutions with a constant background $F^{(4)}$ in M-theory, namely
the AdS$_4\times$S$^7$ and AdS$_7\times$S$^4$ backgrounds --- see,
\eg \cite{duff}. In lifting the D0-branes to M-theory, they become
gravitons carrying momentum in the internal space. Hence the
expanded D2-D0 system considered here correspond to the `giant
gravitons' of ref.~\cite{giant}. The analog of the D2-D0 bound
state in a constant background $F^{(4)}$ corresponds to M2-branes
with internal momentum expanding into AdS$_4$
\cite{goliath,aksun}, while that in a constant $H$ field
corresponds to the M2-branes expanding on S$^4$ \cite{giant}.
Giant gravitons will be discussed at length in the next section.
Alternatively, the dielectric effect has been found to play a role
in other string theory contexts, for example, in the resolution of
certain singularities in the AdS/CFT correspondence \cite{joemat}.
Further, one can consider more sophisticated background field
configurations which through the dielectric effect generate more
complicated noncommutative geometries \cite{sand,shom}. There is
also an interesting generalization to open dielectric branes, in
which the extended brane emerging from the dielectric effect ends
on another D-brane \cite{open}. Other interesting applications of
the dielectric effect for D-branes can be found in
ref.~\cite{hyak}.

\section{Giant Gravitons} \label{giant}

From the above discussion, it seems that in the M-theory
backgrounds of AdS$_{4}\times \S^7$ or AdS$_{7}\times \S^4$, one
will find that an M2-brane carrying internal momentum will expand
into a stable spherical configuration. A Matrix theory description
of such states in terms of noncommutative geometry was only
developed recently \cite{bert}. Instead the original analysis of
these configurations was made in terms of the abelian world-volume
theory of the M2-brane. In fact, the spherical M2-branes expanding
into AdS$_4$ were actually discovered some time ago \cite{oldm}.
It turns out that M5-branes will expand in a similar way for these
backgrounds, and further that expanded D3-branes arise in the type
IIB supergravity background AdS$_{5}\times \S^5$. A detailed
analysis \cite{goliath,giant,aksun} shows that these expanded
branes are BPS states with the quantum numbers of a graviton.
Ref.~\cite{mik} extends this discussion to more general expanded
configurations. In the following, we will discuss the details of
the effect for the D3-branes. Most of the discussion applies
equally well for the analogous M2- and M5-brane configurations.

The line element for $\AdS_5\times \S^5$ may be written as:
\beqa
ds^2&=&-\left(1+{r^2\over L^2}\right)dt^2+{dr^2\over
1+{r^2\over L^2}}+r^2 d\Omega_{3}^2
\labell{metric}\\
&&\qquad\qquad\qquad+
L^2\left(d\theta^2+\cos^2\theta d\phi^2+\sin^2\theta
d\tOmega_{3}^2\right)\ .
\nonumber
\eeqa
This background also involves a self-dual RR five-form field strength
with terms proportional to the volume forms on the two five-dimensional
subspaces: $F^{(5)}= \frac{4}{L}[\varepsilon({\rm AdS}_5)+\varepsilon( \S^5)]$.
With the coordinates chosen above,
the four-form potential on the the AdS part of the space is
\beq
C^{(4)}_{electric}=-{r^{4}\over
L}dt\,\vareps(\S^{3})
\labell{adsp}
\eeq
where $\vareps(\S^{3})$ is the volume form for the
three-sphere described by $d\Omega_{3}^2$. Similarly, the potential
on the $\S^5$ is
\beq
C^{(4)}_{magnetic} = L^{4} \sin^{4}\theta\,d\phi\,\vareps(\tilde{\S}^3)
\labell{sphp}
\eeq
where $\vareps(\tilde{\S}^3)$ is the volume form on $d\tOmega_{3}^2$.
For the D3-brane configurations of interest, the world-volume action
in eqs.~\reef{biact} and \reef{csact} reduces to:
\beq
S_{3}=-T_3\int d^{4}\sigma\ \sqrt{-det(P[G])}+T_3\int P[C^{(4)}]\ .
\labell{actp}
\eeq
Here, the world-volume gauge field has been set to zero, which will
be consistent with the full equations of motion.

Following ref.~\cite{giant}, one can find solutions where a D3-brane
has expanded on the $\S^5$ to a sphere of fixed $\theta$ while it orbits
the $\S^5$ in the $\phi$ direction. Our static gauge choice matches
the spatial world-volume coordinates with the angular coordinates on
$d\tOmega^2_3$, and identifies $\s^0=t$.
Now we consider a trial solution of the form:
$\theta = {\rm constant}$, $r=0$ and $\phi=\phi(t).$
Substituting this ansatz into the world-volume action
\reef{actp} and integrating over the angular coordinates, yields the
following Lagrangian
\beq
\cL_{3}=\frac{\NN}{ L}\left[-\sin^{3}\theta\,\sqrt{1-L^2 \cos^2 \theta
\,\phd^2}+
L\sin^{4} \theta \,\phd\right]\ .
\labell{lag2}
\eeq
Here we have introduced the (large positive) integer $\NN$ which counts
the five-form flux on $\S^5$. This is also, of course, the rank of
the U(N) gauge group in the dual super-Yang-Mills theory.
Introducing the conjugate angular momentum
$P_\phi=\delta\cL_{3}/\delta\phd$, we construct the Hamiltonian:
\beq
\cH_{3}=P_\phi\phd-\cL_{3}
={\NN\over L}\sqrt{p^2+\tan^2\theta\,(p-\sin^{2}\theta)^2}
\labell{ham2}
\eeq
where $p=P_\phi/\NN$.
Given that the Hamiltonian is independent of $\phi$,
the equations of motion will be solved with constant angular momentum (and
hence constant $\phd$). For
fixed $p$, eq.~\reef{ham2} can be regarded as the  potential that determines
the angle $\theta$ for equilibrium. Examining $\cH_{3}$ in detail reveals
degenerate minima at $\sin\theta=0$ and $\sin^2\theta=p$, and
at either of these minima, the energy is $\cH_{3}=\Pp/L$.
The expanded configurations are then the giant gravitons
of ref.~\cite{giant}. An important observation is that
the minima at $\sin^2\theta=p$ only exist
for $p\le1$. As $p$ grows beyond $p=1$, the minima at $\theta\ne0$
are lifted above that at $\sin\theta=0$ and then disappear
completely if $p>9/8$.

The discussion above indicates that one can also consider
the possibility of a brane expanding into the AdS part of the spacetime
\cite{goliath,aksun}.
That is we wish to find solutions where a D3-brane has
expanded to a sphere of constant $r$ while it still orbits in the
$\phi$ direction on the $\S^5$. Choosing static gauge, we  again identify
$\sigma^0= t$ but match the remaining world-volume coordinates
with the angular coordinates on $d\Omega_{3}^2$.
The trial solution is now: $\theta = 0$, $r={\rm constant}$ and
$\phi=\phi(t)$. Beginning with the same\footnote[1]{Our conventions
are such that we actually consider an
anti-D3-brane here \cite{goliath}. That is the  sign of the Wess-Zumino
term in eq.~\reef{actp} was reversed.}
world-volume action \reef{actp},
one calculates as before and arrives at the following Hamiltonian
\beq
\cH_{3}
={\NN\over L}\left[
\sqrt{\left(1+{r^2\over L^2}\right)\left(p^2+{r^{6}\over L^6}
\right)}-{r^{4}\over  L^4}\right]\ .
\labell{green2}
\eeq
where as before $p=\Pp/\NN$.
Examining $\prt \cH_3/\prt r=0$, one finds minima
located at $r=0$ and $\left(r/L\right)^2=p$.
The energy at each of the minima is $\cH_{3}=P_\phi/L$.
In ref.~\cite{goliath}, these expanded configurations were
denoted as dual giant gravitons.
An essential difference from the previous case, however, is that
the minima corresponding to expanded branes persist for arbitrarily large
$p$.

It is interesting to consider the motion of these expanded brane
configurations.
Evaluating $\phd$ for any of the above solutions,
remarkably one finds the same result: $\phd=1/L$, independent of $\Pp$.
Further the center of mass motion for any of the equilibrium
configurations in the full
ten-dimensional background is along a null trajectory. For example,
for the D3-branes expanded on $\S^5$
\beq
ds^2=-(1-L^2\,\cos^2\theta\,\phd^2)\,dt^2=0
\labell{null}
\eeq
when evaluated for $\phd=1/L$ and $\theta=0$(= the center of mass position).
This is, of course, the expected result for a massless `point-like'
graviton, but it applies equally well for both
of the expanded brane configurations.
However, note that in the expanded configurations, the motion
of each element of the sphere is along a time-like trajectory.

From the point of view of five-dimensional supergravity in the
AdS space, the stable brane configurations correspond to massive states with
$M = P_\phi/L$. Their angular momentum means that these states are
also charged under a U(1) subgroup of the SO(6) gauge symmetry in
the reduced supergravity theory. With the appropriate normalizations,
the charge is $Q= P_\phi/L$, and hence one finds that these
configurations satisfy the appropriate BPS bound \cite{giant}.
One can therefore anticipate that all of these configurations should
be supersymmetric. The latter result has been verified by an explicit
analysis of the residual supersymmetries \cite{goliath,aksun}.

The AdS$_5 \times \S^5$ background is a maximally
supersymmetric solution of the type IIB supergravity equations with 32
residual supersymmetries. That is the background fields are invariant under
supersymmetries parameterized by 32 independent Killing spinors. These
Killing spinors are determined by setting
\beq
\delta \Psi_M = {D}_M \eps - \frac{i}{480}\,{\Gamma_M}^{PQRST}
F^{(5)}_{PQRST}\,\eps=0
\eeq
as the variations of all of the other type IIB supergravity fields vanish
automatically. The solutions take the form $\eps=M(x^\mu)\eps_0$
where $\eps_0$ is an arbitrary constant complex Weyl spinor.

A supersymmetric extension of the abelian world-volume action has
been constructed for D3-branes (and all other D$p$-branes) in a
general supergravity background \cite{3brane}. This action can be
viewed as a four-dimensional nonlinear sigma model with a curved
superspace as the target space. Hence the theory is naturally
invariant under the target-space supersymmetry. Further however,
formulating the action with manifest ten-dimensional Lorentz
invariance, requires an additional fermionic invariance on the
world-volume called $\kappa$-symmetry. For a test brane
configuration where both the target space and world-volume
fermions vanish, residual supersymmetries may arise provided there
are Killing spinors which satisfy a combined target-space
supersymmetry and $\kappa$-symmetry transformation. The latter
amounts to imposing a constraint $\Gamma\eps=\eps$ where
\beq
\Gamma=-\frac{i}{4!}\veps^{i_1\cdots i_4} \prt_{i_1}X^{M_1}
\cdots\prt_{i_4}X^{M_4}\Gamma_{M_1\cdots M_4}\ .
\labell{lousya}
\eeq
Of course, this constraint is only evaluated on the D3-brane
world-volume. For all of the minima of the potentials in both
eqs.~\reef{ham2} or \reef{green2},
this constraint reduces to imposing the same projection
\beq
(\Gamma^{t\phi}+1)\eps_0=0\ .
\labell{endresult}
\eeq
Hence not only are the expanded branes and the point-like state
all BPS configurations, all of these configurations preserve
precisely the same supersymmetries. Note that this projection is
what one might have expected for a massless particle moving along
the $\phi$ direction, \eg one can compare this result to the
supersymmetries of gravitational waves propagating in flat space
\cite{wave}.

The brane configurations all preserve one half of the 32
supersymmetries of the background AdS$_5\times\S^5$ spacetime. The
16 supersymmetry transformations satisfying the `wrong'
projection, \ie $(\Ga^{t\phi}-1)\eps_0=0$, would leave the
background spacetime invariant but generate fermionic variations
of the world-volume fields, which at the same time would leave the
energy invariant. Of course, the equations of motion eliminate
half of these to leave 8 fermionic zero-modes in each of the
bosonic configurations studied here. These zero-modes are regarded
as operators acting on a quantum space of states \cite{zero},
which then build up for each bosonic configuration the full
$2^8=256$ states of the supergraviton multiplet, as usual.

Much of the interest in  giant gravitons comes from the suggestion
\cite{giant} that they are related to the `stringy exclusion
principle' \cite{exclus}. The latter arises in the AdS/CFT
correspondence \cite{revue}\ where it is easily understood in the
conformal field theory. A family of chiral primary operators in
the $N$=4 super-Yang-Mills theory terminates at some maximum
weight (and R-charge) because the U(N) gauge group has a finite
rank. In terms of the dual AdS description, these operators are
associated with single particle states and the R-charge of the
operators is dual to angular momentum on the internal five-sphere.
So the appearance of an upper bound on the angular momentum seems
mysterious from the point of view of the supergravity theory,
where there is no apparent upper bound on the Kaluza-Klein
momentum. The suggestion of ref.~\cite{giant} is that if the dual
single particle states are identified with the giant gravitons,
the D3-branes expanded on the $\S^5$. Then the upper bound is
produced by the fact that these BPS states only exist for $p\le1$.
In fact, this exactly reproduces the desired upper bound on the
angular momentum: $P_\phi\le\NN$.

Actually the correct interpretation is slightly more subtle. In
\cite{vijay1}, it was argued that the giant gravitons were dual to
certain subdeterminant operators. While these can be decomposed as
sums of multi-trace operators, the family of operators still
terminates due to the finite rank of the U(N) group with the full
determinant operator. The latter carries R-charge $\NN$. Further
studies \cite{fur1,fur2} have provided strong evidence supporting
this suggestion, at least for large giant gravitons, \ie those
with $P_\phi$ near $\NN$.

Ref.~\cite{aksun} provide some interesting calculations in the
context of the dual CFT. They seem to be able to identify
certain semi-classical field configurations with same properties as
the dual giant gravitons. This suggests a picture where the D3-branes
expanded on AdS$_5$ can be understood in terms of coherent states in the
$N$=4 super-Yang-Mills theory. A complementary description in terms
of large symmetric operators was suggested by \cite{fur1}.

\section{Intersecting Branes} \label{bion}

One interesting aspect of the (abelian) Born-Infeld action
\reef{biact} is that it supports solitonic configurations
describing lower-dimensional branes protruding from the original
D-brane \cite{calmald,gibb,selfd}. For example, in the case of a
D3-brane, one finds spike solutions, known as `bions',
corresponding to fundamental strings and/or D-strings extending
out of the D3-brane. In these configurations, both the
world-volume gauge fields and transverse scalar fields are
excited. The gauge field corresponds to that of a point charge
arising from the end-point of the attached string, {\it i.e.}, an
electric charge for a fundamental string and a magnetic monopole
charge for a  D-string. The scalar field describes the deformation
of the D3-brane geometry caused by attaching the strings. These
solutions seem to have a surprisingly wide range of validity, even
near the core of the spike where the fields are no longer slowly
varying. In fact, one can show that the electric spike
corresponding to a fundamental string is a solution of the full
string theory equations of motion \cite{lars}. Further the
dynamics of these solutions, as probed through small fluctuations,
agrees with the expected string behavior \cite{fluc1}. In part,
these remarkable agreements are probably related to the fact that
these are supersymmetric configurations.

For the system of N D-strings ending on a D3-brane, there is also
a dual description in terms of the nonabelian world-volume theory of
the N D-strings. There one finds solutions which have an interpretation,
in terms of noncommutative geometry, as describing the D-strings expanding
out in a funnel to become an orthogonal D3-brane.
In fact, there is an extensive discussion of this system in the literature
--- see, \eg
\cite{ded,gauntlett} --- where the emphasis was on the close
connection \cite{ded} of the D-string equations to the Nahm
equations for BPS monopoles \cite{nahm}. In ref.~\cite{bion}, our
emphasis was on the interpretation of these solutions in terms of
noncommutative geometry and the remarkable agreement that one
finds with the D3-brane spikes in the large N limit.

For N D-strings in flat space, the dynamics is determined completely by the
Born-Infeld action (\ref{finalbi}) which reduces to \cite{die,ark2}
\begin{equation}
S=-T_1\int d^2\sigma\, \STr\sqrt{-\det\left(\eta_{ab}+
\lambda^2\partial_a\Phi^i Q^{-1}_{ij}\partial_b\Phi^j\right)
%}\sqrt{
\ \det\left(Q^{ij}\right)}\ ,\labell{action2}
\end{equation}
where
\begin{equation}
Q^{ij}=\delta^{ij}+i\lambda[\Phi^i,\Phi^j]\ .
\end{equation}
Implicitly here, the world-volume gauge field has been set to zero,
which will be a consistent truncation for the configurations considered below.
With the usual choice of static gauge, we set
the world-volume coordinates: $\tau=t=x^0$ and $\sigma=x^9$.
For simplicity, one might consider the leading-order (in $\lambda$)
equations of motion coming from this action:
\begin{equation}
\prt^a\prt_a\Phi^i=[\Phi^j,[\Phi^j,\Phi^i]]\ .
\labell{motion}
\end{equation}
Of course, a simple solution of these equations are constant commuting matrices,
as in eq.~\reef{commat}. As mentioned in section \reef{noon}, such
a solution describes N separated parallel D-strings sitting in static
equilibrium.

To find a dual description of the bion solutions of the D3-brane
theory \cite{calmald,gibb}, one needs a static solution which represents the D-strings
expanding into a D3-brane. The corresponding geometry would
be a long funnel where the cross-section at fixed $\sigma$
has the topology of a two-sphere. In this context, the latter cross-section
naturally arises as a fuzzy two-sphere \cite{hoppe,fuzz} if the scalars have
values in an N$\times$N matrix representation of the SU(2) algebra \reef{su2}.
Hence one is lead to consider the ansatz
\begin{equation}
\Phi^i = {R(\sigma)\over\lambda\sqrt{\NN^2-1}}\,\alpha^i,\ \
i=1,2,3,\labell{ansatz}
\end{equation}
where we will focus on case where the $\alpha^i$ are the irreducible $N\times N$
SU(2) matrices. Then with the normalization in eq.~\reef{ansatz}, the function
$|R(\sigma)|$ corresponds precisely to the radius of the fuzzy two-sphere
\begin{equation}
R(\sigma)^2={\lambda^2\over \NN}\sum_{i=1}^3\Tr[\Phi^i(\sigma)^2]\
. \labell{radii}
\end{equation}
Substituting the ansatz \reef{ansatz} into the matrix equations of motion
\reef{motion} yields a single scalar equation
\begin{equation}
R''(\sigma)={8\over\lambda^2(\NN^2-1)} R(\sigma)^3\ ,\labell{leom}
\end{equation}
for which one simple class of solutions is
\begin{equation}
R(\sigma)=\pm {\NN\pi\ls^2\over\sigma-\signot}\left(1-{1\over \NN^2}\right)^{1/2}\ .
\labell{spike}
\end{equation}

Given the above analysis, eqs.~\reef{ansatz} and \reef{spike} only represent a
solution of the leading order equations of motion \reef{motion},
and so naively one expects that it should only be valid for small
radius. However, one can show by direct evaluation
\cite{bion} that in fact these configurations
solve the full equations of motion extremizing the nonabelian action
\reef{action2}. The latter can also be inferred from an analysis
of the world-volume supersymmetry of these configurations. Killing spinor solutions of
the linearized supersymmetry conditions will exist provided
that the scalars satisfy
\begin{equation}
D_\sigma \Phi^i=\pm {i\over2}\varepsilon^{ijk}\left[\Phi^j,
\Phi^k\right]\ .\label{nahmeq}
\end{equation}
The latter can be recognized as the Nahm equations \cite{ded}. Hence
the duality between the D3-brane and D-string descriptions gives a physical
realization of Nahm's transform of the moduli space of BPS magnetic monopoles.
Now inserting the ansatz (\ref{ansatz}) into eq.~\reef{nahmeq} yields
\begin{equation}
R'=\mp {2\over\lambda\sqrt{N^2-1}}R^2\labell{susyeq}
\end{equation}
which one easily verifies is satisfied by the configuration
given in eq.~\reef{spike}. Hence, one concludes
that the solutions given by eqs.~\reef{ansatz} and
\reef{spike} are in fact BPS solutions preserving $1/2$ of the supersymmetry
of the leading order D-string theory. Now in ref.~\cite{aki}, it was shown
that BPS solutions of the leading order theory
are also BPS solutions of the full nonabelian Born-Infeld action
\reef{action2}.

The geometry of the solution, eqs.~\reef{ansatz} and
\reef{spike}, certainly has the desired funnel shape.
The fuzzy two-sphere shrinks to zero size as $\sig\rightarrow\infty$
and opens up to fill the %($x^1$,$x^2$,$x^3$)
$x^{1,2,3}$ hypersurface
at $\sig=\signot$. By examining the nonabelian Wess-Zumino action \reef{finalcs},
one can show that the noncommutative solution induces a coupling to the RR
four-form potential $C^{(4)}_{t123}$. This calculation confirms then that, with
the minus (plus) sign in eq.~\reef{spike},
the D-strings expand into a(n anti-)D3-brane which fills the
%($x^1$,$x^2$,$x^3$)
$x^{1,2,3}$ directions \cite{bion}. Given that the funnel solution of the
D-string theory and the bion spike of the D3-brane theory are both
BPS, one might expect that there will be a good agreement between these
two dual descriptions. The
formula for the height of D3-brane spike above the $x^{1,2,3}$ hyperplane
is \cite{calmald}
\begin{equation}
\sigma-\signot={\NN\pi \ls^2\over R}\ .\labell{reverse}
\end{equation}
Comparing to eq.~\reef{spike}, one finds that for large N the two descriptions
are describing the same geometry up to $1/\NN^2$ corrections. One finds similar
quantitative agreement for large N in calculating the energy, the RR couplings
and the low energy dynamics in the two dual descriptions \cite{bion}.
As in the discussion
of the dielectric effect, one can argue that the D3-brane description is valid
for $R\gg\ls$ while the D-string description is reliable for $R\ll\sqrt{\NN}\ls$
\cite{bion}. Hence one can understand the good agreement between these dual
approaches for large N since there is a large domain of overlap where both are
reliable.

Note that in the configurations considered in this section,
there are no nontrivial supergravity fields in the ambient spacetime.
Hence the appearance of the noncommutative geometry in these solutions
is quite distinct from that in the dielectric effect, where the external
fields drive the D-branes into a certain geometry in the ground state.
In the funnel solutions, the noncommutative geometry was put into the ansatz
\reef{ansatz} by hand. An interesting extension of these solutions is then
to replace the SU(2) generators by those corresponding to some other
noncommutative geometry, \ie to replace eq.~\reef{ansatz} by
\begin{equation}
\Phi^i={R(\sig)\over\lambda\sqrt{C}}\,G^i \labell{newansz}
\end{equation}
where the $G^i$ are new N$\times$N constant matrices satisfying $\sum (G^i)^2 = N\,C$.
An interesting feature of such a construction is that near the
core of the funnel, the leading order equations of motion will still be
those given in eq.~\reef{motion}. Thus for eq.~\reef{newansz} to provide a solution,
the new generators must satisfy $[G^j,[G^j,G^i]]=2a^2\,G^i$ for some constant $a$,
and then the radius is determined by
\begin{equation}
R''= {2a^2\over\lambda^2C} R^3\ , \labell{newdiffer}
\end{equation}
which still has essentially the same form as eq.~\reef{leom} above.
Further the funnel solution of this equation
also has essentially the same form as eq.~\reef{spike}
above, \ie
\begin{equation}
R=\pm{\lambda\sqrt{C}\over a(\sig-\signot)}\ . \labell{newsolll}
\end{equation}
Hence the profile with $R\simeq\lambda/\sig$ is universal for all
funnels on the D-string, independent of the details of the
noncommutative geometry that describes the cross-section of the
funnel.

This universal behavior is curious.
For example, one could consider using this framework to describe a D-string ending
on an orthogonal D$p$-brane with $p>3$. However, from the dual D$p$-brane formulation,
one expects that for large $R$, solutions will
essentially be harmonic functions
behaving like $\sig\propto R^{-(p-2)}$ or $R\propto\sig^{-1/(p-2)}$.
The resolution of this puzzle seems to be that
the two profiles apply in distinct regimes, the first for
small $R$ and the second for large $R$. Hence it must be that the nonlinearity
of the full Born-Infeld action will generate solutions which
display a transition from one kind of behavior to another.

One particular example that we have examined in detail \cite{fiv}
is the case where $G^i$ in eq.~\reef{newansz} are chosen to be
generators describing a fuzzy four-sphere --- these may be found
in, \eg ref.~\cite{wati2}. In this case, the funnel describes the
D-strings expanding into a D5-brane. One does find the expected
transition in the behavior of the geometry. That is,
$\sig\approx\NN^{2/3}\ls/R$ for small $R$ in accord with
eq.~\reef{newsolll}, while at large $R$, higher order terms in the
Born-Infeld action \reef{action2} become important yielding
$\sig\approx\NN^{2/3}\ls^4/R^3$. The same kind of behavior is also
found for the corresponding solutions in the dual D5-brane
world-volume theory, although of course in that case the
nonlinearities of the Born-Infeld action become important for
small $R$. An interesting feature of the D5-brane spike is that it
is also nonabelian in character. Charge conservation arguments
indicate that the D-string acts as a source of the second Chern
class in the world-volume of the D5-brane \cite{semenoff}. More
precisely, if N D-strings end on a collection of D5-branes, then
\begin{equation}
\labell{chargeeq} {1\over 8\pi^2}\int_{\S^4}\Tr(F\wedge F)=\NN\ ,
\end{equation}
for any four-sphere surrounding the D-string endpoint. Hence both
of the dual descriptions have a noncommutative character. Again,
we find that the dual constructions seem to agree at large N,
however, the details of the solutions are more complex \cite{fiv}.
In part, the latter must be due to the fact that the D5$\perp$D1
system is not supersymmetric.

\ack This research was supported in part by NSERC of Canada and
Fonds FCAR du Qu\'ebec. I would like to thank Neil Constable, Marc
Grisaru and \O yvind Tafjord for collaborations in the research
presented in refs.~\cite{goliath,bion,fiv}. I would also like to
thank Mohammad Garousi for his earlier collaborations on the work
appearing in refs.~\cite{garousi1,garousi3}. Finally I would like
to thank the organizers of the Leuven workshop on {\it The quantum
structure of spacetime and the geometrical nature of fundamental
interactions} (September 13--19, 2002) for the opportunity to
speak on the material presented here. I would also like to
congratulate them for organizing a stimulating and successful
meeting in the pleasant surroundings of Leuven.

\appendix

\section{Noncommutative geometry} \label{horse}

The idea that noncommutative geometry should play a role in
physical theories is an old one \cite{old}. Suggestions have been
made that such noncommutative structure may resolve the
ultraviolet divergences of quantum field theories, or appear
in the description of spacetime geometry at the Planck scale.
In the past few years, it has also become a
topic of increasing interest to string theorists. From one point of
view, the essential step in realizing a noncommutative geometry is
replacing the spacetime coordinates by noncommuting operators:
$x^\mu\rightarrow\hx^\mu$. In this replacement, however, there remains
a great deal of freedom in defining the nontrivial commutation relations
which the operators $\hx^\mu$ must satisfy. Some explicit choices that have
appeared in physical problems are as follows:
\vskip 1ex
\noindent\textit{(i) Canonical commutation relations}:
\[
[\hx^\mu,\hx^\nu]=i\theta^{\mu\nu}\qquad \theta^{\mu\nu}\in \mathC
\]
\noindent Such algebras have appeared in the Matrix theory
description of planar D-branes \cite{matrix} --- for a review, see
\cite{watirev}. This work also stimulated an ongoing investigation
by string theorists of noncommutative field theories which arise
in the low energy limit of a planar D-brane with a constant
B-field flux --- see, \eg \cite{seiwit,cds,nikrev}. \vskip 1ex
\noindent\textit{(ii) Quantum space relations}:
\[
\hx^\mu\,\hx^\nu= q^{-1}\, R^{\mu\nu}{}_{\rho\tau}\,\hx^\rho\,\hx^\tau
\qquad R^{\mu\nu}{}_{\rho\tau}\in \mathC
\]
\noindent These algebras received some attention from physicists in the early
1990's --- see, \eg \cite{zum1} --- and have appeared more recently in the geometry
of the moduli space of $N$=4 super-Yang-Mills theory \cite{leigh}.
\vskip 1ex
%\newpage
\noindent\textit{(iii) Lie algebra relations}:
\[
[\hx^\mu,\hx^\nu]=if^{\mu\nu}{}_\rho\,\hx^\rho\qquad f^{\mu\nu}{}_\rho\in
\mathC
\]
\noindent Such algebras naturally arise in the description of
fuzzy spheres as was discovered in early attempts to quantize the
supermembrane \cite{hoppe}. These noncommutative geometries have
also been applied in Matrix theory to describe spherical D-branes
\cite{wati2,wati1}. As discussed in the main text, noncommutative
geometries with a Lie-algebra structure and these noncommutative
spheres, in particular, arise very naturally in various D-brane
systems since the transverse scalars are matrix-valued in the
adjoint representation of U(N). \vskip 1ex

Beginning with a proscribed set of commutation relations for the
coordinates on a given manifold, the bulk of the problem in
noncommutative geometry is to understand the algebra of functions
in this framework. Of course, mathematicians are typically careful
in defining the specific class of functions with which they wish
to work, however, these details are usually glossed over in
physical models. That is, as physicists, we are usually confident
that the physics will guide the choice of functions. For a fuzzy
sphere, one finds that not only is the product structure modified
but that the space of functions is naturally truncated to be
finite dimensional. The remainder of the discussion in the
appendix will focus on these noncommutative spheres
\cite{hoppe,fuzz}, in part because of the emphasis they are given
in the main text. We also elaborate on these examples because, in
contrast to the above discussion, the natural presentation given
below de-emphasizes the role of the commutation relations. In
particular, the fuzzy four-sphere provides an intriguing example
below.

To begin the construction of a fuzzy sphere,
we begin with the standard definition of a $k$-sphere using the embedding
in ($k$+1)-dimensional Cartesian space
\beq
\sum_{i=1}^{k+1}\left(x^i\right)^2=R^2\ ,\qquad\qquad x^i\in\mathR^{k+1}\ .
\labell{sfere}
\eeq
Now functions on $\S^k$ can be expanded in terms of spherical harmonics as
\beq
f(x^i)=\sum_{\ell=0}^\infty f_{i_1\cdots i_\ell} x^{i_1}\cdots x^{i_\ell}
\labell{summ}
\eeq
where $f_{i_1\cdots i_\ell}$ are completely
symmetric and traceless tensors. We could
be more precise in defining a basis of these tensors, but here we will be
satisfied by noting that each
term in the sum is a linear combination of spherical harmonics with principal
quantum number $\ell$. Showing this is straightforward: Denoting the individual
terms in the sum as $f_\ell$ and setting aside the
constraint \reef{sfere}, it is clear that the Laplacian on the Cartesian
space annihilates any of these terms, \ie $\nabla^2 f_\ell =0$.
Now in spherical polar coordinates on $\mathR^{k+1}$, the Laplacian may
be written:
\beq
\nabla^2=R^{-k}\partial_R\left(R^k\partial_R\right)+R^{-2}\nabla^2_\Omega
\labell{laplace}
\eeq
where $\nabla^2_\Omega$ is the angular Laplacian on the unit
$k$-sphere. Hence it follows that $\nabla^2_\Omega
f_\ell=\ell(\ell+k-1)f_\ell$.

In general to produce a fuzzy sphere, one might
proceed by replacing the $k$+1 continuum coordinates above
by finite dimensional matrices, $x^i\rightarrow\hx^i$, whose commutation
relations we leave aside for the moment. The matrices are chosen to
satisfy a constraint analogous to eq.~\reef{sfere}
\beq
\sum_{i=1}^{k+1}\left(\hx^i\right)^2=R^2\,\mathId_\NN\ .
\labell{sfere2}
\eeq
Similarly the continuum functions  are replaced by
\beq
\hat{f}(\hx^i)=\sum_{\ell=0}^{\ell_{max}}
f_{i_1\cdots i_\ell} \hx^{i_1}\cdots \hx^{i_\ell}
\labell{summ2}
\eeq
where $f_{i_1\cdots i_\ell}$ are the same symmetric and traceless
tensors consider in \reef{summ}. Notice that the
`noncommutative' sum is truncated at some $\ell_{max}$ because
for finite dimensional matrices, such products will only yield a
finite number of linearly independent matrices. Thus this matrix
construction truncates the full algebra of functions on the sphere
to those with $\ell\le\ell_{max}$ and the star product on the
fuzzy sphere differs from that obtained by the deformation
quantization of the Poisson structure on the embedding space, \ie
the latter acts on the space of all square integrable functions on
the sphere \cite{kiril}.

The simplest example of this construction is the fuzzy two-sphere, which
was already encountered in sections \ref{diel} and \ref{bion}. In this case,
one chooses $\hx^i=\lambda\,\alpha^i$ with $i=1,2,3$ where the $\alpha^i$
are the generators on the {\it irreducible} N$\times$N representation of SU(2)
satisfying the commutation relations given in eq.~\reef{su2}. These
generators satisfy the Casimir relation
\beq
\sum\left(\alpha^i\right)^2=(\NN^2-1)\,\mathId_\NN
\labell{summ3}
\eeq
and so in order to satisfy the constraint \reef{sfere2}, the normalization
constant should be chosen as
\beq
\lambda={R\over\sqrt{\NN^2-1}}\ .
\labell{umore}
\eeq
With these N$\times$N matrices, one finds the cutoff in eq.~\reef{summ2}
is $\ell_{max}=\NN-1$. So heuristically, we might say that with this
construction we can only resolve distances on the noncommutative sphere
for $\Delta d\gsim R/N$. Hence in the limit $N\rightarrow\infty$, one expects
to get agreement with continuum theory, as was illustrated with the
physical models in the main text.

We should mention that the entire space of functions \reef{summ2}
plays a role in the stringy constructions. To illustrate this point, we
consider the example of the fuzzy two-sphere appearing in the example
of the dielectric effect in section \ref{diel}. In the static ground state
configuration, three of the transverse scalars have a noncommutative
expectation value: $\Phi^i=(f/2)\ \alpha^i$ for $i=1,2,3$. Now one might
consider excitations of this system. In particular, it is natural
to expand fluctuations of the scalars in terms of the
noncommutative spherical harmonics
\beq
\delta\Phi^m(t)=
\psi^m_{i_1...i_{\ell}}(t)\,\alpha^{i_1}
\cdots\alpha^{i_{\ell}}
\labell{expanda1}
\eeq
where as above the coefficients $\psi^m_{i_1\cdots i_{\ell}}$ are completely
symmetric and traceless. In the case of overall transverse scalars,
\ie, $m\ne0,1,2,3$, the linearized equations of motion reduce to
\beqa
\partial^2_t\delta\Phi^m(t)&=&-[\Phi^i,[\Phi^i,\delta\Phi^m(t)]]
\nonumber\\
&=&-\ell(\ell+1)f^2\ \delta\Phi^m(t)\ .
\labell{eomend}
\eeqa
Hence inserting the ansatz $\delta\Phi^m(t)\propto e^{-i\omega t}$, we
find excitations with frequencies $\omega^2=\ell(\ell+1)f^2$. Of course,
these fluctuations inherit the cut off $\ell\le\ell_{max}$ from the
noncommutative framework. The analysis of the fluctuations
in the $i=1,2,3$ directions is more involved but a nice description
is given in \cite{yolk}. In the interesting regime where the dielectric
calculations are expected to be valid, \ie large N and $R\ll\sqrt{\NN}\ls$,
we have $f\ll 1/\sqrt{\NN}\ls$. Hence the low energy excitations are
well below the string scale, giving further corroboration that
the low energy effective action provides an adequate description of the
physics \cite{die}. We might add that these frequencies match the results
found from calculations in the dual continuum framework of the expanded
D2-brane, again up to $1/\NN^2$ corrections. Of course, the latter
description gives no upper cut off on the angular momentum. A similar
discussion \cite{bion} applies for the excitations of the D3$\perp$D1-system
described in section \ref{bion}.

We now turn to a discussion of the fuzzy four-sphere, which is
relevant for the construction of the D-string description of the
D5$\perp$D1 system \cite{fiv}, mentioned at the end of section
\ref{bion}. This construction also played a role in
ref.~\cite{wati2} for the Matrix theory description of spherical
D4-branes (longitudinal M5-branes). At first sight, the
construction of the fuzzy four-sphere appears very similar to the
fuzzy two-sphere above, but in fact it yields a very different
object. One begins by choosing $\hx^i=\lam\, G^i$ where the $G^i$
are an appropriate set of N$\times$N matrices. These matrices were
first constructed in ref.~\cite{grosse} --- see also
ref.~\cite{wati2}. The $G^i$ are given by the totally symmetric
$n$-fold tensor product of $4\times4$ gamma matrices:
\beqa
G^i &=& \left( \Gamma^i \otimes \identity \otimes \cdots
\otimes \identity + \identity \otimes \Gamma^i \otimes \identity
\otimes \cdots \otimes \identity\right.
\labell{gim}\\
&&\left.\qquad\qquad\qquad\qquad\qquad+ \cdots + \identity\otimes
\cdots \otimes\identity \otimes\Gamma^i \right)_{\rm Sym}\ ,
\nonumber \eeqa
where $\Gamma_i$, $i=1,\ldots,5$ are $4 \times 4$ Euclidean gamma
matrices, and $\identity$ is the $4 \times 4$ identity matrix. The
subscript `Sym' means the matrices are restricted to the completely
symmetric tensor product space. With the latter restriction, the
dimension of the matrices becomes
\beq
N={(n+1)(n+2)(n+3)\over6} \labell{Nnrel} \eeq
where $n$ is the integer denoting the size of the tensor product
in eq.~\reef{gim}. The `Casimir' associated with the $G^i$
matrices, \ie $G^i G^i=c\,\identity_\NN$, is given by
\beq c=n(n+4)\ . \labell{kasi} \eeq
Hence to satisfy eq.~\reef{sfere2}, we choose the normalization
constant in $\hx^i=\lam G^i$ as $\lambda=R/\sqrt{n(n+4)}$.

These matrices were presented  as a representation of the fuzzy
four-sphere on the basis of a discussion of representations of
SO(5) in ref.~\cite{grosse}. Working within Matrix theory,
ref.~\cite{wati2} provided a series of physical arguments towards
the same end. That is, the $G^i$ produce a spherical locus, are
rotationally invariant under the action of SO(5) and give an
appropriate spectrum of eigenvalues.

With these $G^i$, one can construct matrix harmonics \reef{summ2}
with $\ell\le\ell_{max}=n$ as before \cite{grosse}. However, a key
difference between the fuzzy two-sphere and the fuzzy four-sphere
is that the $G^i$ do {\it not} form a Lie algebra (in contrast to
the $\al^i$ used to construct the fuzzy two-sphere). As a result
the algebra of these matrix harmonics does not close!
\cite{wati2,grosse} In particular, one finds that the commutators
$G^{ij}\equiv[G^i,G^j]/2$ define linearly independent matrices.
The commutators of the $G^i$ and $G^{jk}$ are easily
obtained\footnote{We refer the interested reader to
refs.~\cite{fiv,wati2} for more details.}:
\begin{eqnarray}
[G^{ij},G^k] &=& 2(\delta^{jk} G^i - \delta^{ik} G^j),
\labell{comme}\cr [G^{ij}, G^{kl}] &=& 2(\delta^{jk}G^{il} +
\delta^{il}G^{jk} - \delta^{ik}G^{jl} - \delta^{jl}G^{ik}).
\nonumber
\end{eqnarray}
Note then that the $G^{ik}$ are the generators of $SO(5)$
rotations. Hence combined the $G^i$ and $G^{ij}$ give a
representation of the algebra $SO(1,5)$, as can be seen from the
definition of the $G^{ij}$ and the commutators in
eq.~\reef{comme}. Hence a closed algebra of matrix functions would
given by
\beq
\tilde{a}_{a_1a_2\cdots a_\ell}\,\tG^{a_1}\tG^{a_2}\cdots
\tG^{a_\ell} \labell{harms3} \eeq
where the $\tG^a$ are generators of $SO(1,5)$ with
$a=1,\ldots,15$, and the $\tilde{a}$ are naturally symmetric in
the $SO(1,5)$ indices. Identifying $\tG^a=G^a$ for $a=1\ldots5$,
the desired matrix harmonics would correspond to the subset of
$\tilde{a}$ with nonvanishing entries only for indices $a_i\le5$.
Thus while the fuzzy four-sphere construction introduces an
algebra that contains a truncated set of the spherical harmonics
on $S^4$, the algebra also contains a large number of elements
transforming under other representations of the $SO(5)$ symmetry
group that acts on the four-sphere. The reader may fine a precise
description of the complete algebra in ref.~\cite{grosse} in terms
of representations of $SO(5)$ (or rather $Spin(5)=Sp(4)$).

Given this extended algebra, or alternatively the appearance of
`spurious' modes, one might question whether the above
construction provides a suitable noncommutative description of the
four-sphere. On the other hand, in the context of nonabelian
D-branes (or Matrix theory), the $G^i$ certainly form the basis
for the description of physically interesting systems such as the
D5$\perp$D1 system \cite{fiv}. In this case, N D-strings open up
into a collection of $n$ perpendicular D5-branes and the fuzzy
four-sphere forms the cross-section of the funnel describing this
geometry. As discussed after eq.~\reef{chargeeq} from the point of
the D5-brane, the four-sphere is endowed with a nontrivial SU($n$)
bundle. In this context, one can show that the relation between N
and $n$ in eq.~\reef{Nnrel} essentially arises from demanding that
the gauge field configuration is homogeneous on the four-sphere
\cite{fiv}. Given that there are extra instantonic or gauge field
degrees of freedom in the full physical system, it is natural that
the `spurious' modes above should be related to non-abelian
excitations of the D5-brane theory. Evidence for this
identification can be found by studying the linearized excitations
of the fuzzy funnel describing the D5$\perp$D1 system and their
couplings to the bulk RR fields \cite{fiv}. A thorough analysis of
the noncommutative geometry \cite{even,kim2} also indicates that
this identification is correct. Hence the fuzzy four-sphere
construction has a natural physical interpretation within string
theory.

\end{document}